\documentclass[times, trackchanges]{aastex63}
\usepackage{inputenc}
\graphicspath{{./}{figures/}}
\usepackage{amsmath}

\newcommand{\mjup}{\ensuremath{M_{\mathrm{Jup}}}}

\newcommand{\Ks}{\mbox{$K_S$\;}}

\begin{document}

\title{Estimating Photometric Distances to Ultracool Dwarfs in Next Generation Space-based Infrared Surveys: Synthetic Photometry and New Absolute Magnitude Versus Spectral Type Relations for JWST, Euclid, and Roman Filters}

\correspondingauthor{Aniket Sanghi}
\email{asanghi@caltech.edu}

\author[0000-0002-1838-4757]{Aniket Sanghi}
\altaffiliation{NSF Graduate Research Fellow.}
\affiliation{Department of Astronomy, California Institute of Technology, 1200 E. California Boulevard, Pasadena, CA 91125, USA}

\author[0000-0003-2232-7664]{Michael C. Liu}
\affiliation{Institute for Astronomy, University of Hawaii, 2680 Woodlawn Drive, Honolulu, HI 96822, USA}

\author[0000-0001-9823-1445]{Trent J. Dupuy}
\affiliation{Institute for Astronomy, University of Edinburgh, Royal Observatory, Blackford Hill, Edinburgh, EH9 3HJ, UK}

\author[0000-0003-0562-1511]{William M. Best}
\affiliation{The University of Texas at Austin, Department of Astronomy, 2515 Speedway, C1400, Austin, TX 78712, USA}

\author[0000-0001-5016-3359]{Robert J. Siverd}
\affiliation{Institute for Astronomy, University of Hawaii, 2680 Woodlawn Drive, Honolulu, HI 96822, USA}

\author[0000-0002-3726-4881]{Zhoujian Zhang}
\affiliation{University of California, Santa Cruz, 1156 High St. Santa Cruz, CA 95064, USA}

\shorttitle{Absolute Magnitude-Spectral Type Relations for JWST, Euclid, and Roman Filters}

\shortauthors{Sanghi et al.}

\begin{abstract}
We synthesize JWST NIRCam photometry for the F164N, F187N, F212N narrow filters, F140M, F162M, F182M, F210M medium filters, and F115W, F150W, F200W wide filters, Euclid Near Infrared Spectrometer and Photometer (NISP) photometry for the $Y_E J_E H_E$ filters, and Roman Wide Field Instrument (WFI) photometry for the F106, F129, F146, F158, F184 and F213 filters using SpeX prism spectra and parallaxes of 688 field-age and 151 young ($\lesssim$ 200 Myr) ultracool dwarfs (spectral types M6--T9). We derive absolute magnitude-spectral type polynomial relations that enable the calculation of photometric distances for ultracool dwarfs observed with JWST, and to be observed with Euclid and Roman, in the absence of parallax measurements. Additionally, using the synthesized photometry to generate color-color figures can help distinguish high-redshift galaxies from brown dwarf interlopers in survey datasets. In particular, anticipating the upcoming Euclid Early Release Observations, we provide synthetic Euclid colors for ultracool dwarfs in our sample.
\end{abstract}

\keywords{Brown dwarfs (185); Photometry (1234); Absolute magnitude (10); Stellar spectral types (2051); James Webb Space Telescope (2291)}

\section{Introduction}
\label{sec:intro}
Ultracool dwarfs are defined as objects with spectral type $\gtrsim$M6 encompassing the lowest-temperature stars as well as brown dwarfs, which are substellar objects more massive than gas-giant planets \cite[$\gtrsim$ 13\;\mjup;][]{2011ApJ...727...57S} but less massive than stars \cite[$\lesssim$ 70~\mjup;][]{2017ApJS..231...15D}. Recently, the unprecedented sensitivity of JWST in the infrared ($\approx$0.6--28.0 $\mu$m) has enabled the detection of previously inaccessible cool brown dwarfs at kiloparsec-scale distances \citep[e.g.,][]{2023ApJ...947L..30C, 2023ApJ...942L..29N, 2023MNRAS.523.4534W, 2024ApJ...962..177B, 2024arXiv240501634G}. Following in JWST's footsteps, in the near future, large scale sky surveys by infrared space telescopes such as Euclid and the Roman Space Telescope are expected to yield a large sample of low-mass substellar objects in the Galactic thick disk and halo that will uniquely probe the structure, formation, and evolution of the Milky Way Galaxy. For such studies, distance is a fundamental parameter. Parallax measurements are generally expensive to obtain for ultracool dwarfs \citep[e.g.][]{2012ApJS..201...19D, 2013AJ....145....2F,  2020AJ....159..257B, 2021AJ....162..102S, 2022ApJ...935...15Z} and will not be available a priori for the new sample of ultracool dwarfs that will be uncovered by the above missions. As an alternative, empirical polynomial relations for absolute magnitude as a function of spectral type can provide inexpensive, albeit less precise, photometric distance estimates. Here, we synthesize JWST NIRCam, Euclid Near Infrared Spectrometer and Photometer (NISP), and Roman Wide Field Instrument (WFI) photometry and provide the corresponding absolute magnitude-spectral type polynomial relations using data from \emph{The UltracoolSheet} version 2.0.0 \citep{best_2024_10573247}. The synthesized photometry provided in this work can also be used to generate color-color figures, which can help distinguish high-redshift galaxies from brown dwarf interlopers \citep[e.g.,][]{2020ApJ...892..125H}.

\section{Synthesizing JWST, Euclid, and Roman Photometry from SpeX Prism Spectra}
For this study, we use a subset of the ultracool dwarf sample presented in \citet{2023ApJ...959...63S}, only including objects with a parallax measurement. This subset is comprised of 688 field-age and 151 young ($\lesssim$ 200 Myr) objects. We refer the reader to Section 2.1 in that paper for a description of the sample. We use flux-calibrated SpeX prism spectra to synthesize JWST NIRCam photometry in a total of 10 photometric bandpasses: F164N, F187N, and F212N narrow bands; F140M, F162M, F182M, F210M medium bands; F115W, F150W, and F200W wide bands. We also synthesize Euclid NISP photometry in the $Y_E J_E H_E$ bandpasses and Roman WFI photometry in the F106, F129, F146, F158, F184 and F213 bandpasses. We used SpeX prism spectra because they have a broad wavelength coverage and capture the near-infrared region (0.8--2.5 $\mu$m) in a single continguous spectrum, thereby avoiding systematic uncertainties associated with combining spectra from differing spectral orders or instruments. The large majority of spectra in our study were obtained from the SpeX Prism Library \citep{2014ASInC..11....7B}. 

We calibrate each object's highest signal-to-noise ratio SpeX spectrum to its observed photometry (PS1 $y$, 2MASS $JH$\Ks, MKO $JHK$, and UKIDSS $JHK$). We follow the procedure described in Section 3.2 of \citet{2023ApJ...959...63S}. Briefly, we retrieve filter transmission profiles for the above photometric bandpasses from the SVO Filter Profile Service\footnote{\url{http://svo2.cab.inta-csic.es/theory/fps/index.php}}. The SpeX spectrum is interpolated using cubic spline interpolation and mapped to the wavelength grid of the transmission profile. We then synthesize photometry in the above bandpasses and compute a scale factor between the observed and synthesized photometry for each bandpass. The final scale factor for the calibration of the SpeX spectrum is obtained as the weighted average of the individual scale factors. Using the flux-calibrated SpeX spectrum and the instrument filter transmission profiles, we synthesize photometry for the photometric bandpasses listed above. The synthesized photometry is publicly available on Zenodo\footnote{\url{https://zenodo.org/records/11199779}}. The apparent magnitudes are converted to absolute magnitudes using parallactic distances. Note that the synthetic photometry (and the following relations) are provided in the Vega magnitude system for JWST NIRCam and in the AB magnitude system for Euclid NISP and Roman WFI.

\begin{figure}
    \centering
    \includegraphics[scale=0.35]{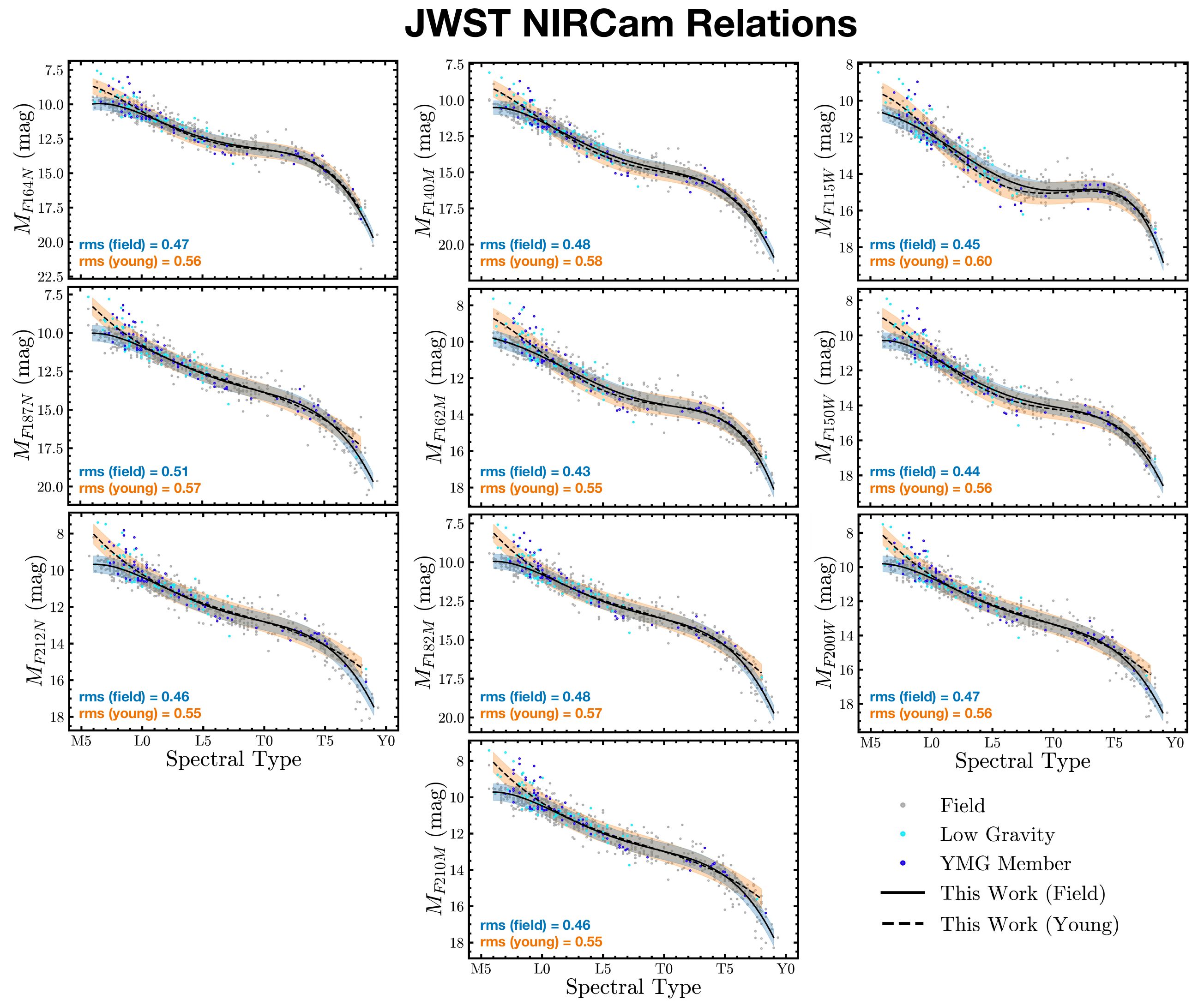}
    \caption{Absolute magnitudes derived for our sample of 1054 ultracool dwarfs as a function of their spectral type (SpT) in various JWST NIRCam filters. Field objects, low gravity objects, and young moving group members are marked by gray, cyan, and blue points respectively. Random scatter in the x direction with amplitude 0.3 SpT is applied to avoid overlapping points. Due to the large number of points, uncertainties in individual measurements are not shown. The median uncertainty in absolute magnitude is 0.055 mag. The black solid and dashed lines are the best-fit polynomial relations for the field and young objects, respectively. The shaded blue and orange regions represent the rms scatter about the field and young object fits, respectively.}
    \label{fig:1}
\end{figure}

\section{Polynomial Relations}
\begin{figure}
    \centering
    \includegraphics[scale=0.26]{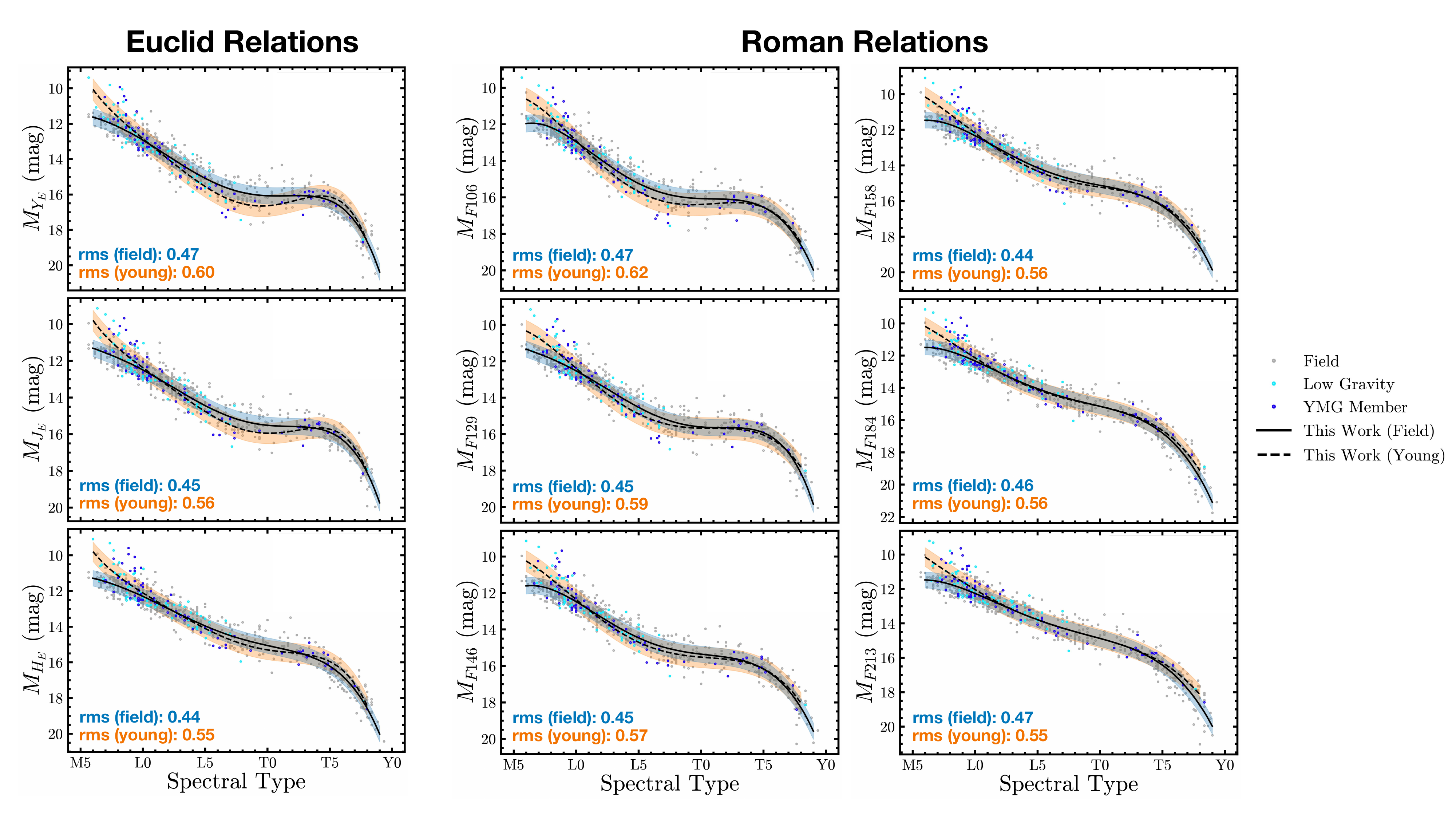}
    \caption{Absolute magnitudes derived for our sample of 1054 ultracool dwarfs as a function of their spectral type (SpT) in various Euclid and Roman filters (AB magnitude system). Field objects, low gravity objects, and young moving group members are marked by gray, cyan, and blue points respectively. Random scatter in the x-direction with amplitude 0.35 SpT is applied to avoid overlapping points. Due to the large number of points, uncertainties in individual measurements are not shown. The median uncertainty in absolute magnitude is 0.045 mag. The black solid and dashed lines are the best-fit polynomial relations for the field and young objects, respectively. The shaded blue and orange regions represent the rms scatter about the field and young object fits, respectively.}
    \label{fig:2}
\end{figure}

We derive empirical relationships between absolute magnitude and spectral type for field and young objects by performing inverse variance-weighted polynomial fits to the data. The polynomial fit is carried out for increasing orders and a lack-of-fit test (\emph{F}-test) is conducted for the best-fit coefficients at each order to determine if a higher-order fit is warranted. We reject the null-hypothesis (higher-order fit required) for $p > 0.05$. We determine a root-mean-square (rms) value to quantitatively characterize the scatter about the best-fit relation. Results are presented in Figures \ref{fig:1} and \ref{fig:2}. The following relations are obtained for field and young objects in the JWST NIRCam filters (Vega magnitude system):

{\footnotesize

\begin{align}
    M_{F164N} &= 16.54 - 2.510\cdot(\mathrm{SpT}) + 0.3153\cdot(\mathrm{SpT})^2 - 0.01465\cdot(\mathrm{SpT})^3 + 0.0002377\cdot(\mathrm{SpT})^4,\;\mathrm{rms}= 0.47,\\
    M_{F187N} &= 14.41 - 1.775\cdot(\mathrm{SpT}) + 0.2316\cdot(\mathrm{SpT})^2 - 0.01065\cdot(\mathrm{SpT})^3 + 0.0001723\cdot(\mathrm{SpT})^4,\;\mathrm{rms}= 0.51,\\
    M_{F212N} &= 13.72 - 1.606\cdot(\mathrm{SpT}) + 0.2069\cdot(\mathrm{SpT})^2 - 0.009491\cdot(\mathrm{SpT})^3 + 0.0001524\cdot(\mathrm{SpT})^4,\;\mathrm{rms}= 0.46,
\end{align}

\begin{align}
    M_{F164N,\; \mathrm{young}} &= 10.94 - 1.410\cdot(\mathrm{SpT}) + 0.2393\cdot(\mathrm{SpT})^2 - 0.01241\cdot(\mathrm{SpT})^3 + 0.0002131\cdot(\mathrm{SpT})^4,\;\mathrm{rms}=0.56,\\
    M_{F187N,\; \mathrm{young}} &= 1.726 + 1.456\cdot(\mathrm{SpT}) - 0.06825\cdot(\mathrm{SpT})^2 + 0.001291\cdot(\mathrm{SpT})^3,\;\mathrm{rms}=0.57,\\
    M_{F212N,\; \mathrm{young}} &= 2.208 + 1.288\cdot(\mathrm{SpT}) - 0.05945\cdot(\mathrm{SpT})^2 + 0.001078\cdot(\mathrm{SpT})^3,\;\mathrm{rms}=0.55,
\end{align}

\begin{align}
    M_{F140M} &= 16.41 - 2.334\cdot(\mathrm{SpT}) + 0.2996\cdot(\mathrm{SpT})^2 - 0.01372\cdot(\mathrm{SpT})^3 + 0.0002187\cdot(\mathrm{SpT})^4,\;\mathrm{rms}=0.48, \\
    M_{F162M} &= 9.333 + 0.1080\cdot(\mathrm{SpT}) - 0.03445\cdot(\mathrm{SpT})^2 + 0.007061\cdot(\mathrm{SpT})^3 - 0.0003882\cdot(\mathrm{SpT})^4 \nonumber \\ 
    & \qquad + 0.000006676\cdot(\mathrm{SpT})^5,\;\mathrm{rms}=0.43,\\
    M_{F182M} &= 14.51 - 1.845\cdot(\mathrm{SpT}) + 0.2411\cdot(\mathrm{SpT})^2 - 0.01120\cdot(\mathrm{SpT})^3 + 0.0001825\cdot(\mathrm{SpT})^4,\;\mathrm{rms}=0.48, \\
    M_{F210M} &= 13.51 - 1.538\cdot(\mathrm{SpT}) + 0.2014\cdot(\mathrm{SpT})^2 - 0.009286\cdot(\mathrm{SpT})^3 + 0.0001497\cdot(\mathrm{SpT})^4,\;\mathrm{rms}=0.46,
\end{align}

\begin{align}
    M_{F140M,\; \mathrm{young}} &= 10.45 - 1.147\cdot(\mathrm{SpT}) + 0.2178\cdot(\mathrm{SpT})^2 - 0.01129\cdot(\mathrm{SpT})^3 + 0.0001915\cdot(\mathrm{SpT})^4,\;\mathrm{rms}=0.58,\\
    M_{F162M,\; \mathrm{young}} &= 9.582 - 0.9275\cdot(\mathrm{SpT}) + 0.1819\cdot(\mathrm{SpT})^2 - 0.009497\cdot(\mathrm{SpT})^3 + 0.0001604\cdot(\mathrm{SpT})^4,\;\mathrm{rms}=0.55,\\
    M_{F182M,\; \mathrm{young}} &= 1.086 + 1.583\cdot(\mathrm{SpT}) - 0.07686\cdot(\mathrm{SpT})^2 + 0.001457\cdot(\mathrm{SpT})^3,\;\mathrm{rms}=0.57,\\
    M_{F210M,\; \mathrm{young}} &= 2.169 + 1.305\cdot(\mathrm{SpT}) - 0.05992\cdot(\mathrm{SpT})^2 + 0.001088\cdot(\mathrm{SpT})^3,\;\mathrm{rms}=0.55,
\end{align}

\begin{align}
    M_{F115W} &= 8.705 + 0.7676\cdot(\mathrm{SpT}) - 0.1525\cdot(\mathrm{SpT})^2 + 0.01752\cdot(\mathrm{SpT})^3 - 0.0008005\cdot(\mathrm{SpT})^4 \nonumber \\ 
    & \qquad + 0.00001243\cdot(\mathrm{SpT})^5,\;\mathrm{rms}=0.45,\\
    M_{F150W} &= 15.63 - 2.113\cdot(\mathrm{SpT}) + 0.2714\cdot(\mathrm{SpT})^2 - 0.01238\cdot(\mathrm{SpT})^3 + 0.0001951\cdot(\mathrm{SpT})^4,\;\mathrm{rms}=0.44,\\
    M_{F200W} &= 13.54 - 1.542\cdot(\mathrm{SpT}) + 0.2046\cdot(\mathrm{SpT})^2 - 0.009471\cdot(\mathrm{SpT})^3 + 0.0001536\cdot(\mathrm{SpT})^4,\;\mathrm{rms}=0.47,
\end{align}

\begin{align}
    M_{F115W,\; \mathrm{young}} &= 13.42 - 1.994\cdot(\mathrm{SpT}) + 0.3101\cdot(\mathrm{SpT})^2 - 0.01517\cdot(\mathrm{SpT})^3 + 0.0002427\cdot(\mathrm{SpT})^4,\;\mathrm{rms}=0.60,\\
    M_{F150W,\; \mathrm{young}} &= 9.638 - 0.9020\cdot(\mathrm{SpT}) + 0.1848\cdot(\mathrm{SpT})^2 - 0.009671\cdot(\mathrm{SpT})^3 + 0.0001628\cdot(\mathrm{SpT})^4,\;\mathrm{rms}=0.56,\\
    M_{F200W,\; \mathrm{young}} &= 1.888 + 1.385\cdot(\mathrm{SpT}) - 0.06433\cdot(\mathrm{SpT})^2 + 0.001187\cdot(\mathrm{SpT})^3,\;\mathrm{rms}=0.56.
\end{align}
}

\noindent
The following relations are obtained for field and young objects in the Euclid filters (AB magnitude system):

{\footnotesize

\begin{align}
    M_{Y_E} &= 10.77 + 0.3068\cdot(\mathrm{SpT}) - 0.08364\cdot(\mathrm{SpT})^2 + 0.01302\cdot(\mathrm{SpT})^3 - 0.0006665\cdot(\mathrm{SpT})^4 + 0.00001096\cdot(\mathrm{SpT})^5,\;\mathrm{rms}= 0.47,\\
    M_{J_E} &= 8.361 + 1.137\cdot(\mathrm{SpT}) - 0.1983\cdot(\mathrm{SpT})^2 + 0.01977\cdot(\mathrm{SpT})^3 - 0.0008425\cdot(\mathrm{SpT})^4 + 0.00001263\cdot(\mathrm{SpT})^5,\;\mathrm{rms}= 0.45,\\
    M_{H_E} &= 10.96 + 0.02483\cdot(\mathrm{SpT}) - 0.01845\cdot(\mathrm{SpT})^2 + 0.005700\cdot(\mathrm{SpT})^3 - 0.0003353\cdot(\mathrm{SpT})^4 + 0.000005948\cdot(\mathrm{SpT})^5,\;\mathrm{rms}= 0.44,
\end{align}

\begin{align}
    M_{Y_E,\; \mathrm{young}} &= -4.836 + 5.220\cdot(\mathrm{SpT}) - 0.7132\cdot(\mathrm{SpT})^2 + 0.05355\cdot(\mathrm{SpT})^3 - 0.001939\cdot(\mathrm{SpT})^4 + 0.00002631\cdot(\mathrm{SpT})^5,\;\mathrm{rms}= 0.60,\\
    M_{J_E,\; \mathrm{young}} &= -6.369 + 5.719\cdot(\mathrm{SpT}) - 0.7760\cdot(\mathrm{SpT})^2 + 0.05620\cdot(\mathrm{SpT})^3 - 0.001964\cdot(\mathrm{SpT})^4 + 0.00002590\cdot(\mathrm{SpT})^5,\;\mathrm{rms}= 0.56,\\
    M_{H_E,\; \mathrm{young}} &= -3.138 + 4.385\cdot(\mathrm{SpT}) - 0.5643\cdot(\mathrm{SpT})^2 + 0.03970\cdot(\mathrm{SpT})^3 - 0.001368\cdot(\mathrm{SpT})^4 + 0.00001805\cdot(\mathrm{SpT})^5,\;\mathrm{rms}= 0.55.
\end{align}
}

\noindent
We obtain the following relations for field and young objects in the Roman filters (AB magnitude system):

{\footnotesize
\begin{align}
    M_{F106} &= 20.05 - 3.097\cdot(\mathrm{SpT}) + 0.3870\cdot(\mathrm{SpT})^2 - 0.01752\cdot(\mathrm{SpT})^3 - 0.0002711\cdot(\mathrm{SpT})^4,\;\mathrm{rms}= 0.47,\\
    M_{F129} &= 8.268 + 1.204\cdot(\mathrm{SpT}) - 0.2131\cdot(\mathrm{SpT})^2 + 0.02119\cdot(\mathrm{SpT})^3 - 0.0009002\cdot(\mathrm{SpT})^4 + 0.00001345\cdot(\mathrm{SpT})^5,\;\mathrm{rms}= 0.45,\\
    M_{F146} &= 18.25 - 2.555\cdot(\mathrm{SpT}) + 0.3202\cdot(\mathrm{SpT})^2 - 0.01451\cdot(\mathrm{SpT})^3 + 0.0002262\cdot(\mathrm{SpT})^4,\;\mathrm{rms}= 0.45, \\
    M_{F158} &= 16.60 - 2.037\cdot(\mathrm{SpT}) + 0.2621\cdot(\mathrm{SpT})^2 - 0.01201\cdot(\mathrm{SpT})^3 + 0.0001905\cdot(\mathrm{SpT})^4,\;\mathrm{rms}= 0.44,\\
    M_{F184} &= 16.23 - 1.893\cdot(\mathrm{SpT}) + 0.2460\cdot(\mathrm{SpT})^2 - 0.01141\cdot(\mathrm{SpT})^3 + 0.0001856\cdot(\mathrm{SpT})^4,\;\mathrm{rms}= 0.46,\\
    M_{F213} &= 14.96 - 1.434\cdot(\mathrm{SpT}) + 0.1899\cdot(\mathrm{SpT})^2 - 0.008770\cdot(\mathrm{SpT})^3 + 0.0001425\cdot(\mathrm{SpT})^4,\;\mathrm{rms}= 0.47,
\end{align}

\begin{align}
    M_{F106,\; \mathrm{young}} &= 14.34 - 2.048\cdot(\mathrm{SpT}) + 0.3249\cdot(\mathrm{SpT})^2 - 0.01603\cdot(\mathrm{SpT})^3 + 0.0002582\cdot(\mathrm{SpT})^4,\;\mathrm{rms}= 0.62,\\
    M_{F129,\; \mathrm{young}} &= 13.19 - 1.661\cdot(\mathrm{SpT}) + 0.2700\cdot(\mathrm{SpT})^2 - 0.01334\cdot(\mathrm{SpT})^3 + 0.0002153\cdot(\mathrm{SpT})^4,\;\mathrm{rms}= 0.59,\\
    M_{F146,\; \mathrm{young}} &= 11.50 - 1.135\cdot(\mathrm{SpT}) + 0.2140\cdot(\mathrm{SpT})^2 - 0.01101\cdot(\mathrm{SpT})^3 + 0.0001827\cdot(\mathrm{SpT})^4,\;\mathrm{rms}= 0.57, \\
    M_{F158,\; \mathrm{young}} &= 10.43 - 0.7689\cdot(\mathrm{SpT}) + 0.1694\cdot(\mathrm{SpT})^2 - 0.009046\cdot(\mathrm{SpT})^3 + 0.0001548\cdot(\mathrm{SpT})^4,\;\mathrm{rms}= 0.56,\\
    M_{F184,\; \mathrm{young}} &= 9.684 - 0.4830\cdot(\mathrm{SpT}) + 0.1350\cdot(\mathrm{SpT})^2 - 0.007565\cdot(\mathrm{SpT})^3 + 0.0001354\cdot(\mathrm{SpT})^4,\;\mathrm{rms}= 0.56,\\
    M_{F213,\; \mathrm{young}} &= 7.799 + 0.1890\cdot(\mathrm{SpT}) + 0.05307\cdot(\mathrm{SpT})^2 - 0.003680\cdot(\mathrm{SpT})^3 + 0.00007194\cdot(\mathrm{SpT})^4,\;\mathrm{rms}= 0.55.
\end{align}
}

In all of the above relations, $\mathrm{SpT}$ is a numerical spectral type: M6 = 6, L0 = 10, L5 = 15, T0 = 20, T5 = 25, etc. For the field relations $6 \le \mathrm{SpT} \le 29$ and for the young relations $6 \le \mathrm{SpT} \le 28$. We use optical spectral types for M and L dwarfs when available and infrared spectral types otherwise. We use infrared spectral types for T and Y dwarfs. 

\section{Color-Color Selection of Ultracool Dwarfs with Euclid}
\begin{figure}
    \centering
    \includegraphics[scale=0.3]{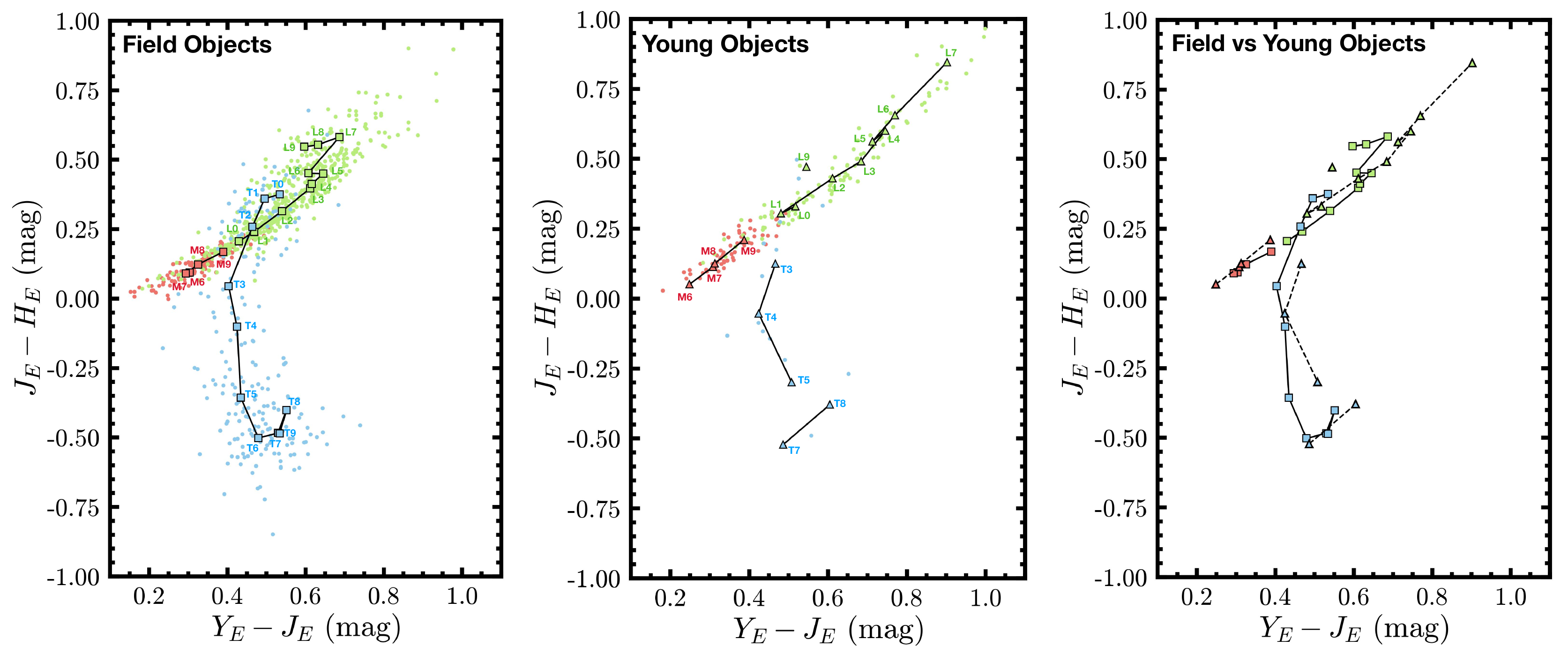}
    \caption{Color-color diagram of field (left) and young (center) ultracool dwarfs in our parallax-based sample in the Euclid filters. A comparison of the mean colors is presented in the right panel. For both field and young objects, synthesized colors for individual objects are shown as red, green, and blue circles corresponding to M, L, and T spectral types. The average color computed across each spectral bin is marked by a square and triangle for field and young objects respectively. The lines connecting the squares and triangles are for representative purposes and do not correspond to any particular fit to the data.}
    \label{fig:3}
\end{figure}

In anticipation of Euclid's upcoming Early Release Observations, we provide color information in the Euclid filters for ultracool dwarfs in our sample as a function of spectral type. This will enable the photometric identification of new candidate ultracool dwarfs. Using the synthesized Euclid photometry, we computed $Y_E - J_E$ and $J_E - H_E$ colors for spectral types M6-T9 for both field and young objects in our sample. The characteristic color as a function of spectral type is computed as the average and the uncertainty as the standard deviation of the colors of all objects in our sample in a spectral type bin. We do not have young objects in our sample for spectral types L8, T0, T1, T2, T6, T9. Note that there are only a small number of young T dwarfs in our sample and the colors thus may not be representative of the population. Colors are presented in Table \ref{table:1} and Figure \ref{fig:3}. 

\begin{deluxetable*}{ccccccc}
    \label{table:1}
    \centering
    \tablecaption{Synthethic Euclid colors for ultracool dwarfs in our sample as a function of spectral type.}
    \tablehead{\colhead{Spectral Type} & \colhead{$(Y_E - J_E)_{\mathrm{field}}$} & \colhead{$(J_E - H_E)_{\mathrm{field}}$}  & \colhead{$N_{\mathrm{obj, field}}$} &  \colhead{$(Y_E - J_E)_{\mathrm{young}} $} & \colhead{$(J_E - H_E)_{\mathrm{young}}$} & \colhead{$N_{\mathrm{obj, young}}$} \\ \colhead{} & \colhead{(AB mag)} & \colhead{(AB mag)} & \colhead{} & \colhead{(AB mag)} & \colhead{(AB mag)} & \colhead{}}
    \startdata
M6 & $0.304 \pm 0.044$ & $0.093 \pm 0.044$ & 5 & 0.249 & 0.051 & 1 \\ 
M7 & $0.294 \pm 0.062$ & $0.090 \pm 0.043$ & 21 & $0.308 \pm 0.030$ & $0.113 \pm 0.043$ & 3 \\ 
M8 & $0.325 \pm 0.060$ & $0.122 \pm 0.042$ & 45 & $0.313 \pm 0.061$ & $0.125 \pm 0.032$ & 8 \\ 
M9 & $0.390 \pm 0.055$ & $0.168 \pm 0.038$ & 35 & $0.387 \pm 0.045$ & $0.210 \pm 0.057$ & 19 \\ 
L0 & $0.430 \pm 0.085$ & $0.206 \pm 0.072$ & 72 & $0.518 \pm 0.105$ & $0.330 \pm 0.100$ & 26 \\ 
L1 & $0.468 \pm 0.091$ & $0.240 \pm 0.079$ & 71 & $0.481 \pm 0.104$ & $0.305 \pm 0.087$ & 9 \\ 
L2 & $0.540 \pm 0.079$ & $0.314 \pm 0.080$ & 53 & $0.612 \pm 0.138$ & $0.430 \pm 0.141$ & 9 \\ 
L3 & $0.611 \pm 0.091$ & $0.396 \pm 0.095$ & 34 & $0.684 \pm 0.105$ & $0.491 \pm 0.132$ & 11 \\ 
L4 & $0.616 \pm 0.077$ & $0.412 \pm 0.081$ & 36 & $0.746 \pm 0.106$ & $0.600 \pm 0.127$ & 12 \\ 
L5 & $0.645 \pm 0.111$ & $0.449 \pm 0.108$ & 37 & $0.713 \pm 0.143$ & $0.562 \pm 0.172$ & 12 \\ 
L6 & $0.607 \pm 0.128$ & $0.451 \pm 0.128$ & 24 & $0.770 \pm 0.212$ & $0.655 \pm 0.266$ & 4 \\ 
L7 & $0.687 \pm 0.131$ & $0.581 \pm 0.137$ & 19 & $0.902 \pm 0.102$ & $0.845 \pm 0.159$ & 8 \\ 
L8 & $0.632 \pm 0.075$ & $0.553 \pm 0.093$ & 25 & \nodata & \nodata & 0 \\ 
L9 & $0.597 \pm 0.076$ & $0.546 \pm 0.103$ & 24 & 0.545 & 0.471 & 1 \\ 
T0 & $0.534 \pm 0.058$ & $0.375 \pm 0.098$ & 14 & \nodata & \nodata & 0 \\ 
T1 & $0.495 \pm 0.073$ & $0.359 \pm 0.064$ & 12 & \nodata & \nodata & 0 \\ 
T2 & $0.464 \pm 0.060$ & $0.258 \pm 0.052$ & 16 & \nodata & \nodata & 0 \\ 
T3 & $0.403 \pm 0.106$ & $0.044 \pm 0.100$ & 10 & $0.467 \pm 0.121$ & $0.125 \pm 0.236$ & 3 \\ 
T4 & $0.425 \pm 0.060$ & $-0.102 \pm 0.083$ & 8 & 0.424 & -0.054 & 1 \\ 
T5 & $0.435 \pm 0.071$ & $-0.357 \pm 0.082$ & 23 & 0.508 & -0.300 & 1 \\ 
T6 & $0.480 \pm 0.048$ & $-0.503 \pm 0.122$ & 18 & \nodata & \nodata & 0 \\ 
T7 & $0.530 \pm 0.069$ & $-0.484 \pm 0.070$ & 19 & 0.487 & -0.522 & 1 \\ 
T8 & $0.552 \pm 0.096$ & $-0.402 \pm 0.504$ & 15 & $0.605 \pm 0.067$ & $-0.380 \pm 0.156$ & 2 \\ 
T9 & $0.534 \pm 0.035$ & $-0.486 \pm 0.026$ & 2 & \nodata & \nodata & 0 \\ 
\enddata
\tablecomments{$N_{\mathrm{obj}}$ designates the number of objects in a given spectral type bin using which the average color and uncertainty were computed. We present the color value without uncertainties for spectral types where only one object was available since a standard deviation cannot be calculated.}
\end{deluxetable*}

\section*{Acknowledgments}
This work has benefitted from \emph{The UltracoolSheet}, maintained by Will Best, Trent Dupuy, Michael Liu, Aniket Sanghi, Rob Siverd, and Zhoujian Zhang, and developed from compilations by \citet{2012ApJS..201...19D}, \citet{2013Sci...341.1492D}, \citet{2016ApJ...833...96L}, \citet{2018ApJS..234....1B}, \citet{2020AJ....159..257B}, \citet{2023ApJ...959...63S}, and \citet{2023AJ....166..103S}. 

\bibliography{\string references}

\end{document}